\documentclass[preprint,superscriptaddress,preprintnumbers,pre,showpacs]{revtex4}
\usepackage{amsmath}
\usepackage{graphicx}

\newcommand\be{\begin{equation}}
\newcommand\ee{\end{equation}}                                                                                
\newcommand\ba{\begin{eqnarray}}
\newcommand\ea{\end{eqnarray}}

\begin{document}

\title{
  Mid-infrared femtosecond laser pulse filamentation in
  hollow waveguides: a comparison of simulation methods
}

\author{J. Andreasen}
\affiliation{
  College of Optical Sciences, University of Arizona,
  Tucson, AZ 85721, U.S.A.
}
\author{M. Kolesik}
\affiliation{
  College of Optical Sciences, University of Arizona,
  Tucson, AZ 85721, U.S.A.
}
\affiliation{
  Department of Physics, Constantine the Philosopher University, 
  Nitra 94974, Slovakia
}

\date{\today}

\begin{abstract}
This work compares computational methods for laser pulse propagation in hollow waveguides filled with rare gases 
at high pressures, with applications in extreme nonlinear optics in the mid-infrared wavelength region.
As the wavelength of light $\lambda=2\pi/k$ increases with respect to the transverse size $R$ of a leaky waveguide,
the loss of light out of the waveguide upon propagation, in general, increases.
The now standard numerical approach for studying such structures is based on expansion of the propagating field
into approximate leaky waveguide modes. 
We compare this approach to a new method that resolves the electric field in real space and correctly captures the 
energy loss through the waveguide wall. 
The comparison reveals that the expansion-based approach overestimates losses that occur in 
nonlinearly reshaped pulsed waveforms. 
For a modest increase in computational effort, the new method offers a physically more accurate model to describe 
phenomena (e.g., extreme pulse-selfcompression) in waveguides with smaller values of $kR$.

\end{abstract}

\maketitle

\section{Introduction}

The focus of this work is on simulation methods for extreme nonlinear regimes of laser pulse
propagation and optical filamentation~\cite{Couairon2007,BergeRPP07,chin_femtosecond_2009} within hollow
waveguides filled with gas, which can be under high pressure.  
There are two particularly important applications of filamentation confined in such lossy structures to which
the methods discussed here apply.
In the first, high-power femtosecond pulses propagate and give rise to filaments in a quasi-planar waveguide
geometry made of two glass blocks spaced apart by a few hundred micrometers. 
The groups of Midorikawa~\cite{nurhuda_optical_2006,chen_compression_2008} and Mysyrowicz~\cite{Akturk:09,Arnold:09} 
investigated such slab-waveguide arrangements for pulses in the near infrared region
as a means of improving power-scaling in femtosecond filamentation, while partially suppressing unwanted break-up 
into multiple, small-scale filamentary structures. 
They also demonstrated that efficient pulse self-compression can be achieved due to the nonlinear response of gas
and propagation effects induced by the leaky wave-guiding structure. 

Closely related systems are capillaries utilized for pulse self-compression and/or high-harmonic generation.
One especially promising application in high-harmonic generation uses a highly pressurized capillary with 
an inert gas as a table-top source of coherent XUV radiation.  
As demonstrated recently by Popmintchev and co-authors~\cite{popmintchevS12}, a novel regime occurs in the 
generation of extremely high harmonic orders (up to five thousand). 
Numerical simulations suggest that the favorable dynamics is the result of an interplay between filamentation, 
with its underlying intensity clamping by the generated plasma, and effects of the waveguide that suppress
higher order transverse propagation modes.

Both categories, i.e., slab and hollow capillary waveguides, are usually simulated using essentially the same 
approach. 
Building on the resonant ``metastable''  modes of the wave-guiding structure, it expands the optical field 
into a set of orthogonal basis functions that approximate the actual modal fields. 
The leaky nature of the modes is accounted for by adding a suitable imaginary
part to the propagation constant of each basis function. 
However, as the wavelength of light $\lambda=2\pi/k$ increases with respect to the transverse system size $R$, 
the loss increases and this approximation breaks down.
Therefore, it can be difficult to accurately simulate systems with smaller values of $kR$.
Currently, high-intensity light-matter interactions in ultra-short laser pulses are being explored at
wavelengths up to 4 $\mu$m (see e.g. \cite{popmintchevS12,silva_multi-octave_2012}), and it is expected 
that even longer-wavelength regimes will be studied in the near future. 
We therefore concentrate on the region beyond 4 $\mu$m.

The purpose of this paper is to present two computational 
methods that bypass the need for the lossy-mode approximation.
The first one is based on the recently developed \cite{gUPPE} generalized Unidirectional Pulse Propagation 
Equations (gUPPE), which treats the total field in real space and incorporates loss implicitly through 
Maxwell's equations.
This is utilized mainly for testing purposes of the second method discussed here, which is a further approximation
applied to gUPPE such that the simulation domain only encompasses the bore of the waveguide.
This approximation reduces both the total grid size of the simulation and the computational cost associated with
simulating fine field features which can manifest in the cladding.
The approach then becomes similar to the standard method in efficiency,
but with the advantage that the losses of light into the capillary cladding are treated in a more realistic way. 
A consequence of a more accurate treatment of loss is the precision to which nonlinear effects in the waveguide,
which intimately depend on the local light intensity, can be modeled.

\section{Approximate leaky modes expansion: Summary}

The standard simulation technique is briefly discussed here.
For the purposes of reference in this paper, we term this method ALMEX, 
standing for Approximate Leaky-Mode EXpansion. 
Its main features are:

\begin{itemize}

\item[$\bullet$]
The electromagnetic field outside of the waveguide bore is
neglected, and a scalar approximation is adopted so that
the linearly polarized field is considered axially symmetric
and confined to a domain with radius $R$.
This approximation is based on the fact that 
lower-order resonant modes of the lossy waveguide exhibit
long ``lifetimes,'' and have very small amplitudes in the
cladding.

\item[$\bullet$]
Imaginary part of the resonant modal function is neglected. 
This is justified due to its relatively small amplitude in comparison with the real part, 
and due to its characteristic shape: it vanishes in the center of the hollow waveguide 
and only increases significantly close to the cladding.

\item[$\bullet$]
Real part of the resonant modal function, in the case of a capillary, 
is approximated by a Bessel mode of an idealized circular waveguide with a perfectly electric conducting shell. 
Unlike the real resonant modes, the approximated functions are all
zero at the edge of the radial computational domain:
\begin{equation*}
\phantom{MM}{\cal M}_n = J_0( k_n r) \  ,  \  J_0(k_n R) = 0 \  , \  n=1,2,\ldots,N_\perp \ ,
\end{equation*}
and the boundary condition selects the spectrum of allowed transverse wavenumbers $k_n$.

\item[$\bullet$]
The propagation constant of the above Bessel-like mode is modified by 
the addition of an imaginary part that is adjusted such that 
each mode exhibits the same linear loss upon propagation
as the corresponding leaky mode:
\begin{equation}
\phantom{MM}\beta_n  = \sqrt{k(\omega)^2 - k_n^2 }   + i \alpha_n  \  , \ k(\omega) = \frac{\omega n_g(\omega)}{c},
\end{equation}
where $n_g$ is the refractive index of the gas pressurized in the hollow waveguide and
\begin{equation}
\alpha_n = \left(\frac{u_n}{2 \pi}\right)^2 \frac{\lambda^2}{a^3}\frac{1 + n_{cl}^2}{2 \sqrt{n_{cl}^2-1}} 
\end{equation}
is the imaginary part of the leaky mode propagation constant~\cite{Marcatili}; 
$u_n$ is the $n$-th zero of $J_0$.

\item[$\bullet$]
Spectral propagator based on these approximate modes is then included in
a nonlinear laser pulse propagation simulator. Here we utilize the 
Unidirectional Pulse Propagation Equations (UPPE)
which is well suited for this purpose because it is a fully spectral solver~\cite{KolesikPRE04}.
The spectral propagator for axially symmetric situations is implemented
with the help of a Hankel transform, and the waveform is ``natively''
expanded in the Bessel modes. The sole modification of the
free-space algorithm therefore consists in modifying modal propagation
constants as described above.

\end{itemize}

Implementation of this approach has been described in detail, for example,
in Ref. ~\cite{Arnold:09}, in the context of planar waveguides. 
Reference~\cite{popmintchevS12} provides many simulation-related details, including 
the nonlinear medium model, for the case of a pressurized capillary waveguide.

\section{Generalized UPPE: Summary}

The generalized Unidirectional Pulse Propagation Equations (gUPPE)
were designed \cite{gUPPE} for structures characterized by a frequency-dependent relative
permittivity dependent on the transverse coordinates,
$\epsilon= \epsilon(\vec r_\perp,\omega)$ .
The framework takes advantage of the previous UPPE algorithm~\cite{KolesikPRE04}, and
marries with it beam propagation techniques suitable for 
problems with material contrasts and sharp interfaces 
parallel to the propagation direction. 

We recount here the primary equations involved for ease of reference.
The linear propagator in this method is ``generated'' by 
the operator $\hat L$ which is  related to the corresponding 
Helmholtz equation. It acts on the transverse electric vector field as
\be
 \hat L \vec E_\perp \equiv   
  \frac{\omega^2}{c^2} \epsilon(r_\perp, \omega)\vec E_\perp  
+ \Delta_\perp \vec E_\perp
+ \nabla \frac{1}{\epsilon}\vec E_\perp . \nabla_\perp \epsilon .
\label{eq:linearprop}
\ee
All other light-matter interactions, in particular nonlinearity, are included in the 
operator $\hat N$
\be
\hat N[ \vec E ] \equiv 
\frac{\omega^2}{\epsilon_0 c^2} \vec P(\vec E)
+ \nabla \frac{1}{\epsilon_0\epsilon} \nabla \cdot \vec P(\vec E) \ ,
\label{eq:N}
\ee
where $\vec P(\vec E)$ encapsulates the model of the nonlinear medium response.
The algorithmic separation of the nonlinear medium from the linear beam-propagation 
aspects of the problem can be formulated exactly, but for practical calculations one
needs to invoke the so-called uni-directional approximation: If nonlinear interactions
are weak enough so that they fail to generate backward propagating waves, the
system can be restricted to the forward-propagating field alone,
\be
E^F_\perp = e^{+i \sqrt{\hat L} z} A^F_\perp (z) \ ,
\label{eq:EvsA}
\ee
in which the spectral amplitudes $A^F$ are the native variables evolved by the solver
according to the equation
\be
\partial_z A^F_\perp(r_\perp,\omega,z) = + \frac{i}{2\sqrt{\hat L}} e^{-i \sqrt{\hat L} z} \hat N_\perp[ e^{+i \sqrt{\hat L} z}  A^F] \ .
\label{eq:general}
\ee
Various beam-propagation techniques can be utilized to implement the
the linear propagator $\exp(-i \sqrt{\hat L} z)$. In the present work, we
use a Pad\`e approximant in the spirit of a multistep method
akin to that by Hadley~\cite{hadley_wide-angle_1992,hadley_multistep_1992}
\be
e^{i \sqrt{\hat L} \Delta z}=e^{i \beta \sqrt{1 + \hat X} \Delta z} = \prod_k \frac{\hat X + a_k}{\hat X + b_k},
\ee
where $\beta^2(\omega) \equiv \omega^2\epsilon(\omega) / c^2$ stands for the dominant part of the Helmholtz operator, 
and the coefficients $a_k,b_k$ depend on $\Delta z$ and are chosen as to reproduce the Taylor 
expansion of the left hand side. The following specialization is used in both gUPPE-based methods
discussed:
\be
\frac{4 i + (i - \beta \Delta z)\hat X }{4 i + (i + \beta \Delta z)\hat X } e^{i\beta \Delta z} \ .
\label{eq:padeapp}
\ee
Having specified an implementation for the linear propagator, the numerical technique to solve gUPPE 
reduces to a solution of a very large system of ordinary differential equations.
Thanks to the diagonal nature of the linear propagator in angular frequency space, its numerical application 
is equivalent to a set of uncoupled beam-propagation problems, and can be efficiently parallelized. 

Details of the derivation can be found in Ref.~\cite{gUPPE}, and we refer the reader to Ref.~\cite{Guide11}
for an in-depth description of numerical methods suitable for the UPPE-type ODE system.

\section{$\rm g$UPPE for a hollow waveguide}

As an alternative to the ALMEX method, gUPPE can be used to simulate ultrashort pulses propagating through 
pressurized hollow waveguides. 
Propagation losses in capillaries increase steeply for smaller radii,
and that is why typical diameters in experimental applications are several hundred micrometers. 
That results in a propagation regime that is paraxial for the
field propagating within the bore of the waveguide. 
Also, the polarization of the inner beam is dominated by a single transverse component, 
while the longitudinal component can be safely neglected. 

These are essentially the assumptions that underline the ALMEX method discussed earlier. 
In contrast, the main feature of gUPPE in the context of waveguides is:

\begin{itemize}

\item[$\bullet$]
Fields outside of the waveguide, propagating through the capillary ``cladding'' glass, 
are resolved in the simulation.
All leakage characteristics of such a regime is implicitly included.

\item[$\bullet$]
The computational domain includes a perfectly matched layer
which absorbs the radiation, and thus models a capillary shell with
an infinite extent.

\end{itemize}

The physical difference between ALMEX and gUPPE is in the loss mechanism. 
By resolving the whole relevant domain including the capillary cladding, gUPPE can
properly model the dynamics of light leaking from the bore into the cladding.  
While the overall loss rate is reasonably well captured by ALMEX, gUPPE can properly reflect the fact
that ``instantaneous'' losses by beams that are localized away from the gas-glass interface are much 
lower than for beam configurations which are in contact with the interface. 

\section{$\rm g$UPPE with lossy boundary conditions}

This Section describes an approximation in which a lossy boundary
condition is applied at the bore-cladding interface $r=R_{max}$,
and the optical field is only sampled withing the bore of the waveguide.
In short, we refer to the resulting method as gUPPE-b.
This approach offers savings in computational effort in more than one way. 
First, the required computational domain is about one half of that required to include cladding 
plus a perfectly matched layer absorbing boundary condition.
Second, the grid resolution can be coarser, because only the field profiles within the waveguide are
simulated, and they are paraxial. 
Fields in the higher-index cladding obtain small transverse wavelengths and require finer resolution.
Third, the integration step in the propagation direction
can be longer when the method has no need to properly resolve the light propagating in the cladding.

The approximation presented here is designed 
for any waveguiding regime that is characterized by relatively low losses or leakage of the light
into the cladding. 
This means that the transverse wavenumbers of waves comprising the laser pulse field are small in 
comparison to those of light that propagate through the cladding.
Thus, we can estimate the angle of propagation in the cladding by that which corresponds to a grazing
incidence angle, $\sin(\alpha) \sim 1/n_{cl}$. 
We will use this approximation to construct the boundary condition for the field at $r=R_{max}$ as follows.

To simplify the notation, let us assume that the refractive index of gas in the capillary is equal to one. 
Consider the spectral component of the field with angular frequency $\omega$ and the dispersion relations
inside the waveguide and cladding
\begin{equation}
k_{i}^2 + k_\parallel^2 = \frac{\omega^2}{c^2} \ \ \ , \ \ \
k_{o}^2 + k_\parallel^2 = \frac{\omega^2 n_{cl}^2}{c^2} \ \ \ , \ \ \
\end{equation}
where $k_i$ and $k_o$ are the transverse wavenumbers in the gas and in the glass, respectively, and $k_\parallel$ is
the longitudinal wavenumber. 
The latter must be the same on both sides of the interface due to field continuity conditions.
As pointed out above, in the regime we are interested in,
$k_i$ is small in comparison to $k_o$ and to the terms on the right hand side of the dispersion relation
(this is nothing but a statement of the paraxial approximation).
Neglecting $k_i$ altogether, we can estimate
\begin{equation}
k_o = \frac{\omega}{c} \sqrt{n_{cl}^2 -1 } \ ,
\end{equation}
which will be used in the boundary condition.
Imposing  continuity of the field derivative with respect to the direction normal to the interface $\partial_x$, 
we can write
\begin{equation}
  \partial_x E_{i} = \partial_x E_{o} = i k_o E_{o} =  i E_{o}\frac{\omega}{c} \sqrt{n_{cl}^2 -1 },
\end{equation}
with $E_{i}$ and $E_{o}$ standing for the field value at the interface approached
from inside and outside, respectively.
For the derivative on the left, we use a three-point estimate
\begin{equation}
\partial_x E_{i} \approx \frac{3}{2}\frac{E_{i} - E_{1}}{\Delta x} -\frac{1}{2}\frac{E_{1} - E_{2}}{\Delta x},
\end{equation}
where  $E_{1}$ and $E_{2}$ are samples of the field at points a distance $\Delta x$ and $2 \Delta x$
from the interface. 
For TE and TM polarization we have the conditions
\begin{equation}
E_{o} = E_{i} \ \ \ {\rm and} \ \ \ n_{cl}^2 E_{o} = E_{i} \  ,
\end{equation}
respectively, and $E_{i}$ is obtained as
\begin{align}
  E_{i}^{TM} =& \frac{4E_1 - E_2}{3 - 2ik_{o}\Delta x/n_{cl}^2} \nonumber\\
  E_{i}^{TE} =& \frac{4E_1 - E_2}{3 - 2ik_{o}\Delta x}. \label{eq:abc1}
\end{align}

The ALMEX method approximates the EH$_{1n}$ mode, which is dominated by a linearly polarized field.
The boundary condition at $y=0$, $x=\pm R_{max}$ is therefore of the TE type, while it is of TM type
for $y=\pm R_{max}$, $x=0$, and is ``mixed'' elsewhere along the interface. 
Because gUPPE utilizes the BPM propagator in a purely linear regime, the TE and TM boundary conditions are  
averaged so that the solution represents the axially symmetric component of the electric field:
\begin{equation}
E_{i} = \frac{1}{2}(E_{i}^{TM} + E_{i}^{TE}) \ .
\label{eq:abc2}
\end{equation}
It can be shown that the resulting boundary reflectivity is smaller than one
for all nonzero transverse wavenumbers, and that consequently, its implementation
should result in a stable propagation algorithm.
With the boundary condition approximated by Eqs. (\ref{eq:abc1}) and (\ref{eq:abc2}),
the computational grid is restricted to the inside of the waveguide, and the
operator $\hat X$ in Eq. (\ref{eq:padeapp}) reduces to a radial Laplacian.  
The action of the linear propagator is then calculated in a scheme closely resembling a Crank-Nicolson-type step.

\section{Method validation}

To validate the gUPPE-b method and, in particular, its lossy boundary condition
at the interface between gas and capillary glass, we investigate the propagation
of a femtosecond pulse in a linear regime. 
The central wavelength of the pulse is chosen to be 6~$\mu$m, and the refractive index of the 
capillary glass 1.3.
The pulse is focused to a 100~$\mu$m beam radius ($1/e^2$) at the entrance of
a capillary with radius 150~$\mu$m. 
This input condition excites some higher order modes and results in propagation that exhibits beating
as it gradually tends toward the lowest order mode. 

\begin{figure}[ht]
\scalebox{0.5}{\includegraphics[clip]{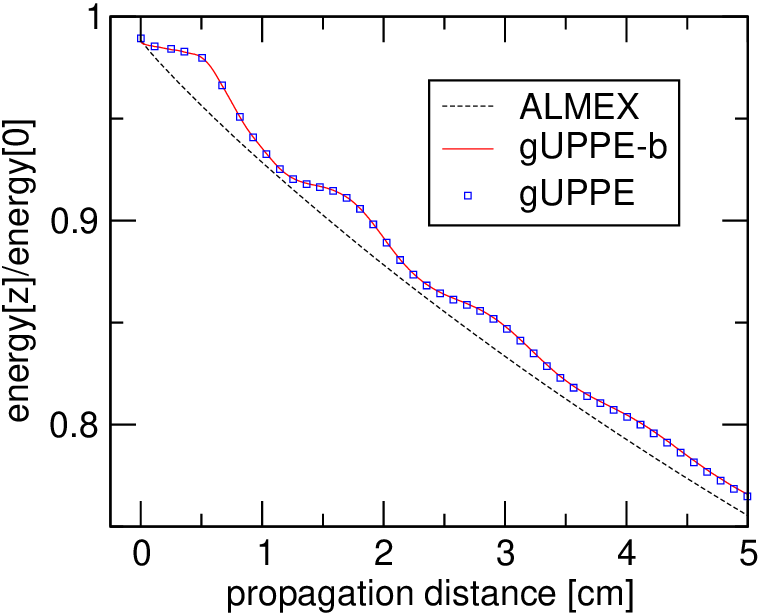}}\\
\scalebox{0.5}{\includegraphics[clip]{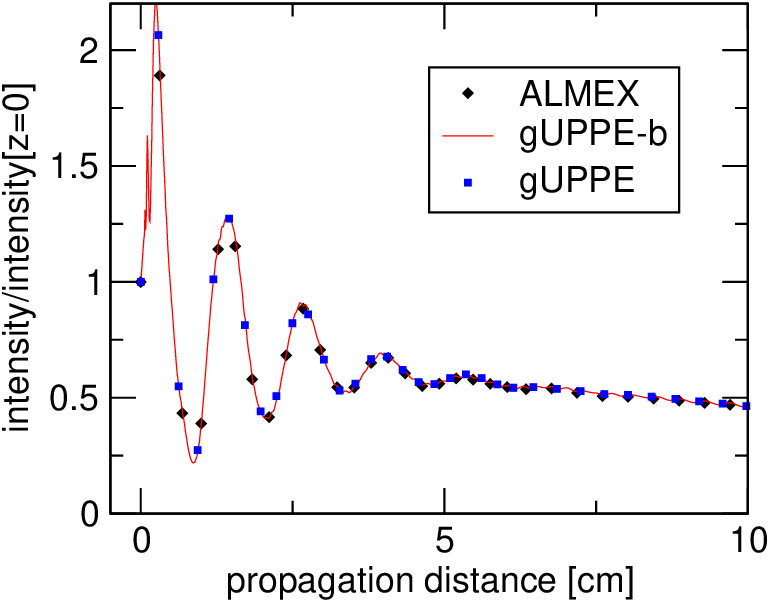}}
\caption{
Comparison of three different methods in a linear propagation
regime for a pulse focused at the entrance
of the capillary waveguide. Top panel: evolution
of the pulse energy reveals how instantaneous losses
reflect that a beam localized away from the capillary wall
suffers lower loss. This effect is not captured by the ALMEX method. 
Lower panel depicts the evolution of the maximum intensity.
}
\label{fig:comp-ABR}
\end{figure}

Figure~\ref{fig:comp-ABR} shows results from all three methods. 
The upper panel depicts the pulse energy within the capillary hollow bore as a function of propagation distance
for the first few centimeters. 
Note that there is very little energy loss in both gUPPE methods over the first few millimeters. 
This is because the beam diameter is still too small to reach the gas-glass interface. 
Similarly, other ``nearly-horizontal'' sections of the plot arise when the beam is localized away 
from the interface and cannot lose energy (which only occurs through the interface). 
On the other hand, the energy loss in the ALMEX method only reflects relative weights of higher-order
modes (which suffer higher losses) but not their relative phases. 
As such, the resulting instantaneous loss is insensitive to whether or not the 
wavepacket is actually in contact with the capillary interface. 
This is, of course, an artifact caused by approximations that underline ALMEX. 
While the overall energy loss rate averaged over several centimeters of propagation is indeed captured
correctly in the ALMEX approach, both gUPPE methods correctly reflect the
dynamics of the light leaking into the capillary shell.

The lower panel in Fig.~\ref{fig:comp-ABR} shows that all three methods agree
quite well in terms of the maximum on-axis intensity. 
This measure is less sensitive to the physical difference between ALMEX and gUPPE, 
at least in the linear regime. 

The comparison of the two gUPPE-based methods shows that their mutual agreement is excellent, 
which in turn indicates that the lossy boundary condition utilized in gUPPE-b is satisfactory. 
Having validated this new method, it will be utilized in the rest of this work. 
It is significantly faster than gUPPE, with its speed comparable to that of ALMEX.
The next Section is to demonstrate its utility for the simulation of
extremely nonlinear regimes in long-wavelength laser pulse dynamics.

\section{Comparative simulations of confined filamentation in mid-infrared pulses}

The aim of this Section is to demonstrate that while the now standard ALMEX approach can give
qualitatively valid results, the gUPPE-b method with lossy boundary conditions provides better treatment of the 
laser pulse interaction with the capillary. 
The improvement is achieved without significant computational cost increase, and gUPPE-b should therefore 
be the preferred method of choice, especially for longer wavelengths, where loss is greater.

The choice of examples for our comparative simulations has been motivated by increasing interest in 
near- and mid-infrared wavelength laser pulses and their nonlinear interactions with media, such as
supercontinuum or high-harmonic generation.

\begin{figure}[t]
\scalebox{0.5}{\includegraphics[clip]{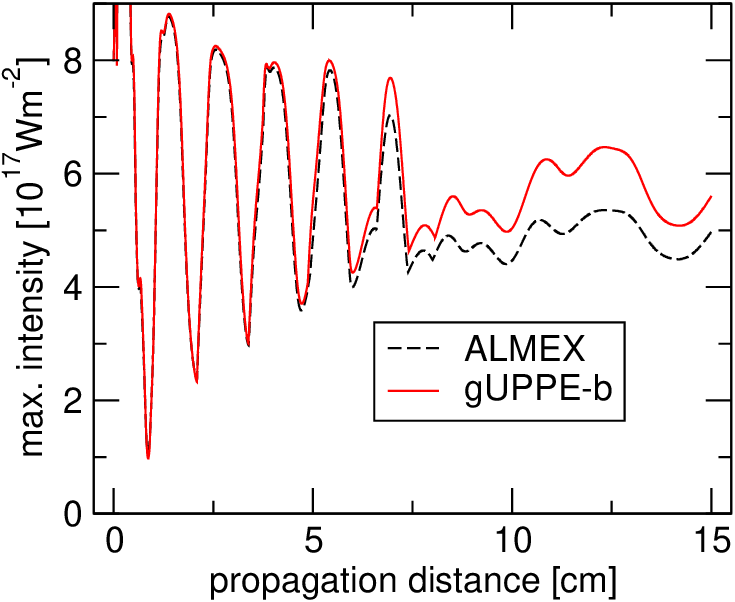}}\\
\scalebox{0.5}{\includegraphics[clip]{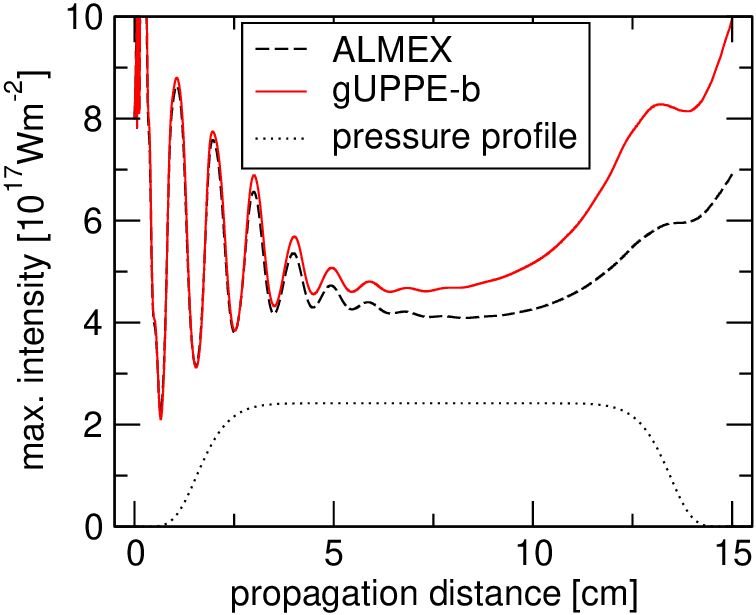}}
\caption{
  Maximal on-axis intensity produced by initially 100-fs duration
  pulses at 6~$\mu$m (top) and 8~$\mu$m (bottom) wavelengths, propagating 
  through a hollow waveguide filled with Argon at 20~atm. 
}
\label{fig:inteAC}
\end{figure}

We consider 100~fs laser pulses with central wavelengths of 6 or 8~$\mu$m,
focused on the entrance into a capillary with inner radius 150~$\mu$m.
We assume that the capillary is pressurized to 20~atm with Argon through
small openings located a few centimeters from both ends, and that a flat-top pressure 
profile exists within the capillary. 
Because computational issues are our main concern in this work, we assume that the laser 
pulse at the input is a simple collimated Gaussian pulse with a $1/e^2$ intensity radius of 100~$\mu$m. 
Note that to obtain results closer to the experiment,
propagation of the laser pulse before the capillary entrance should also be considered~\cite{popmintchevS12}.

We have chosen the same input pulse energy for all simulations presented.
The pulse energy, the length of the hollow waveguide (15~cm), and its inner radius,
were all selected as to achieve optimal self-compression at the end of the pressurized region for the
longer wavelength of 8~$\mu$m. 
For the sake of comparison, we keep parameters unchanged when we switch to a 6~$\mu$m wavelength and/or 
to a different simulation method.

The refractive index of the capillary glass was chosen as 1.3, and also kept unchanged
for simulations at both wavelengths. 
For gUPPE-b, the radial grid is equidistant, using 75 samples, while 30 samples were used in the spectral, 
discrete Hankel transform-based ALMEX method. 
In both cases, doubling these numbers does not lead to visible changes in the observed simulated quantities.
Temporal grids spanned 4~ps, and had 8192 sampling points. 
3100 active (angular) frequencies were used in the bandwidth between 
$1.2\times 10^{14}$ s$^{-1}$ and $5.0\times 10^{15}$ s$^{-1}$.
Typical running times were 1700 and 2700 sec in ALMEX and gUPPE-b simulations, respectively. 

Figure~\ref{fig:inteAC} reports simulated on-axis maximum intensities for the two methods at both wavelengths. 
In both cases, the ALMEX method underestimates the peak intensity, with the difference growing along the
propagation distance. 
Having seen that the same quantity was captured equally well in the linear regime, 
we conclude that the difference here is due to nonlinear interactions with the gas. 
This has the tendency to induce higher-order modes into the propagation, which mainly occurs where the intensity
is high, i.e., in the center of the waveguide. 
The ALMEX method overestimates the losses imposed on these nonlinear waveforms. 
For the gUPPE-b method, losses are effectively lower in the presence of nonlinear interactions
because the new waves tend to be localized in the center, limiting the loss through
contact with the capillary cladding interface.
Thus, ALMEX, in effect, adds certain artificial damping to nonlinear interactions.

\begin{figure}[ht]
\scalebox{0.5}{\includegraphics[clip]{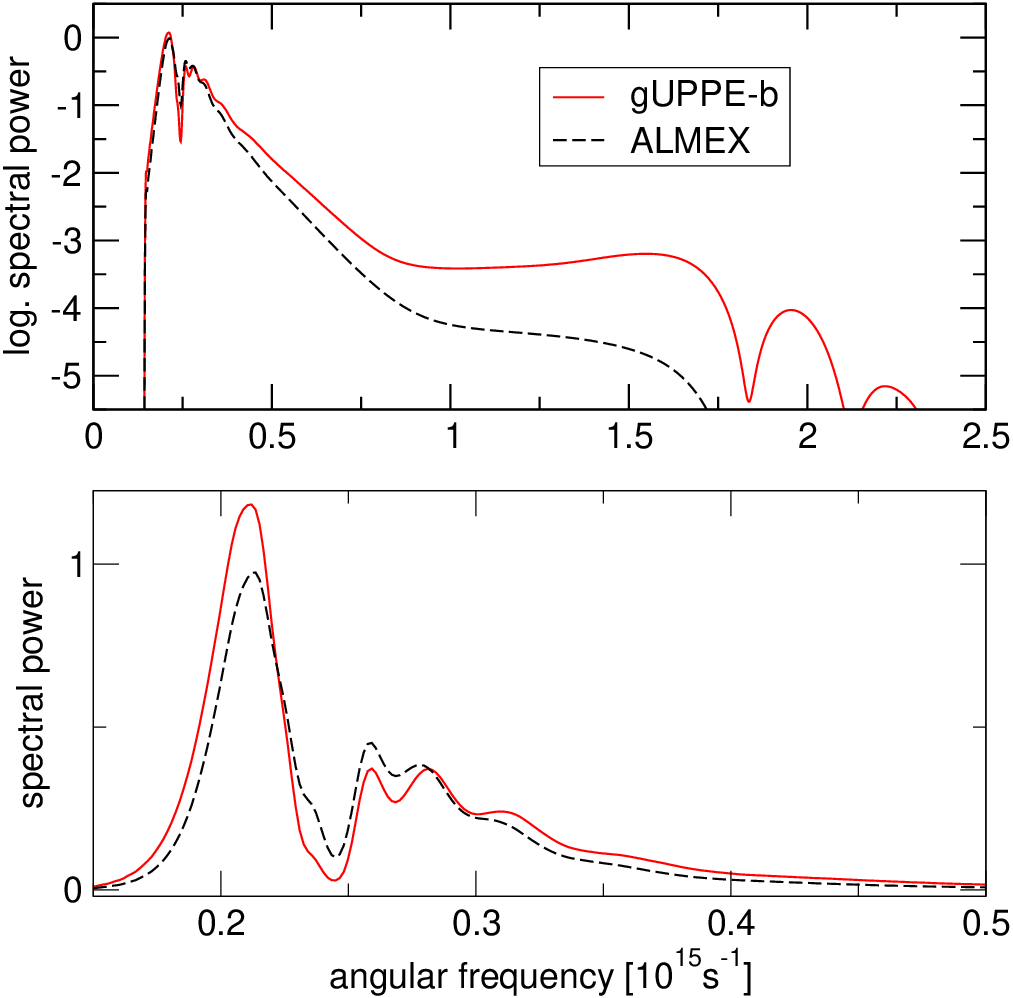}}
\caption{
  Supercontinuum generation in a hollow waveguide. 
  The excitation wavelength of the 100~fs-duration pulse is 8~$\mu$m. 
  Stronger supercontinuum generation in the gUPPE-b method reflects sharper pulse self-compression.
  The linear-scale spectrum in the bottom panel compares the central portion of the spectra.
}
\label{fig:spcAC}
\end{figure}

The high intensities that occur along axis of the hollow waveguide
are the result of mainly temporal reshaping of the pulsed waveforms.
In particular, sharp trailing edges develop in the second
half of the capillary, and these manifest as broad
spectral supercontinua shown in Fig.~\ref{fig:spcAC}. 
In line with the result on intensities, the ALMEX method exhibits weaker spectral broadening,
especially far from the initial excitation wavelength.

\begin{figure}[ht]
\scalebox{0.45}{\includegraphics[clip]{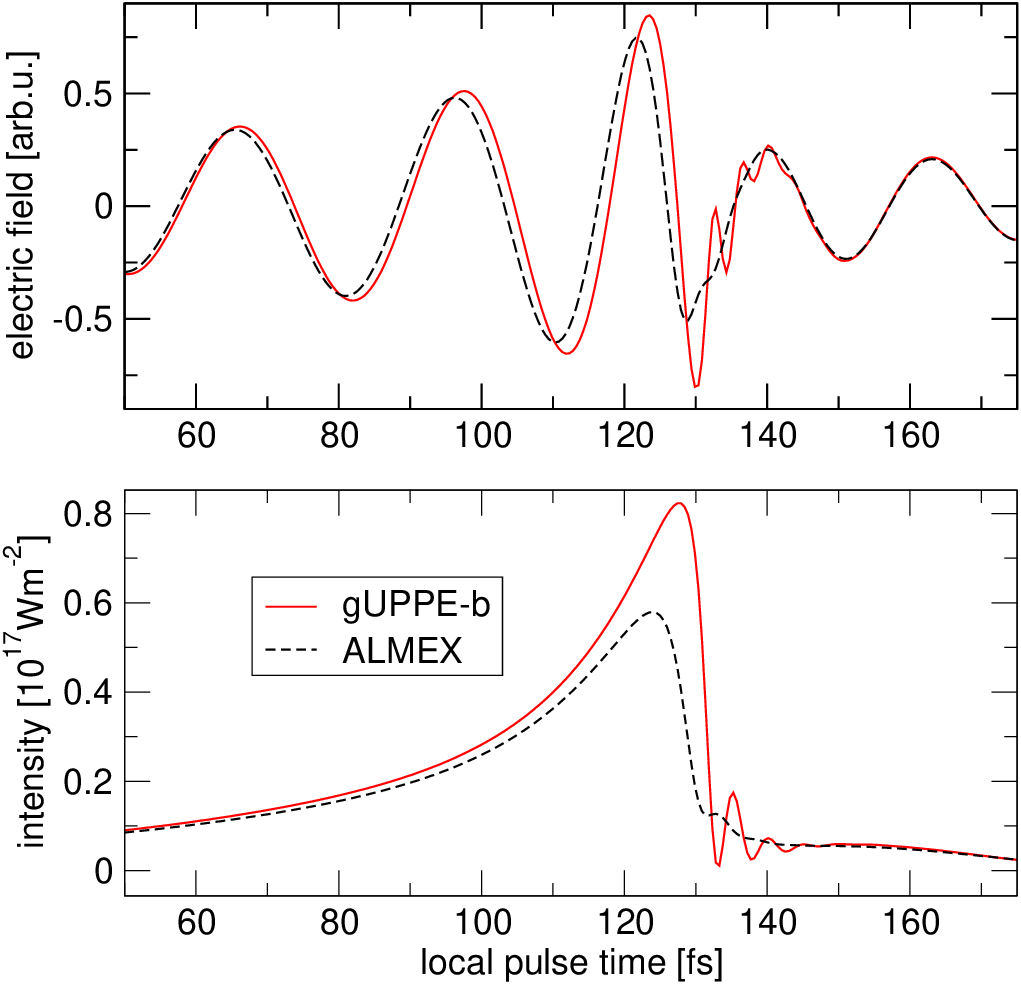}}
\caption{
  Pulse self-compression  within a pressurized hollow waveguide
  for an initially 100~fs pulse at $\lambda=8 \mu$m.
  Top: Electric field waveforms obtained in different methods tend to agree in 
  the leading edge of the pulse, and deviate around the peak. 
  The gUPPE-b method also shows stronger chirp, and high-frequency components 
  ``localized'' in the trailing   portion of the pulse. 
  Bottom: Cycle-averaged intensity.
}
\label{fig:pulsesAC}
\end{figure}

Figure~\ref{fig:pulsesAC} depicts intensity temporal profiles that develop at the propagation distance 
of 12~cm, which is at the end of the high-pressure section of the hollow waveguide.
We have chosen the parameters as to achieve as strong self-compression as possible for $\lambda = 8$~$\mu$m. 
The bottom panel indeed shows that the originally 100~fs pulse develops a high-intensity, few-cycle feature. 
Interestingly, as one can infer from the intensity versus propagation distance plot, 
this feature is rather robust and survives for several centimeters. 
Besides the steep trailing edge, another typical characteristic of these waveforms is extreme chirp,
when the ``local wavelength'' shortens significantly within a single cycle. 
These pulses exhibit a high ratio between the peak electric field values in the 
strongest and next-to-strongest sub-cycles. 
Therefore, it should be interesting to investigate the dynamics of high-harmonic generation in these pulses,
especially the influence of rapidly changing frequency --- this will be addressed elsewhere.

From the point of view of method comparison, Fig.~\ref{fig:pulsesAC}
essentially confirms our findings on intensity and spectra;
the ALMEX approach gives results that qualitatively follow the
more accurate gUPPE-b method, but it exhibits lower resulting intensities.
Based on experience from a number of comparative simulation runs, we point out that, in general, 
the agreement between the two methods is rather close in the leading edge of the pulsed waveform. 
Deviations appear in the middle and trailing parts of the pulse.  
Physically, it is the trailing edge of the pulse where waves are reflected from the
capillary wall, and it is these waves that behave differently in the two approaches. 
In gUPPE-b, nonlinear features localized in the center need to diffract and reach the capillary wall
before they experience significant losses. 
On the other hand, the losses in ALMEX set in immediately and only reflect the 
spatial power spectrum of the beam.

Among others, application of the gUPPE-b method is envisioned in the field
of high-harmonic generation (HHG). 
Spatial and temporal evolution of the driving pulses are of crucial importance for efficient 
sourcing of harmonic generation, especially in the filamentation regime~\cite{popmintchevS12}.
A detailed investigation of HHG dynamics in these peculiar waveforms will be involved and, therefore,
discussed in a dedicated work.
Based on preliminary results, we merely mention here the importance of 
accurate modeling of spatial and temporal evolution for long-wavelength HHG-driving pulses.

\section{Summary\label{sec:sum}}

We have introduced a new method to simulate high-intensity 
mid-infrared laser pulses propagating in hollow waveguides.
The method is based on the generalized unidirectional pulse propagation framework~\cite{gUPPE} 
into which is included an approximate lossy boundary condition that represents the interaction 
between the pulse and the hollow waveguide inner wall. 

We have compared the new method, gUPPE-b, to the currently standard approach for simulating
pulse propagation in leaky waveguides, ALMEX, which builds on expansion of the
optical field into approximate lossy modes of the given hollow waveguide.
Both gUPPE-b and ALMEX treat the electric field of the propagating pulse in semi-vectorial approximation, 
utilize the same light-mater interaction model, and are both implemented using the same software package 
for a large system of ordinary differential equations. 
The difference between the methods lies in how the interaction with the hollow waveguide is treated. 
gUPPE-b uses a local semi-transparent boundary that mimics the waveguide wall and causes the loss of light
through leakage into the waveguide cladding. 
In contrast, ALMEX approximates these losses by modification of propagation constants
of its orthogonal  modal expansion.

We first validated the new method gUPPE-b against fully spatially resolved simulations (gUPPE)
including the capillary glass. 
The subsequent comparative simulations show that at long wavelengths, 
gUPPE-b should be the preferred method over ALMEX. 
It is somewhat slower, but it can correctly capture the dynamic nature of losses and their 
dependence on the localization of the beam within the hollow waveguide.
In ALMEX, nonlinear waveforms suffer additional unphysical loss and, as a consequence,
the simulated waveforms attain maximum intensities up to 25\% lower than those simulated by gUPPE-b. 
The simulated waveforms also differ in the amount of frequency chirp they acquire upon 
propagation --- the chirp can be very strong in the regime favorable for pulse self-compression, 
when the local frequency can change significantly within a single half-cycle.

Besides the hollow capillary waveguides, the proposed method should be useful for 
laser pulse propagation in generic leaky waveguide systems, for example,
in filamentation regimes confined to the semi-planar geometry in slab waveguides.

\begin{acknowledgments}
  This work was supported by the U.S. Air Force Office for Scientific
  Research under grants FA9550-11-1-0144, and FA9550-10-1-0561.
\end{acknowledgments}

\end{document}